\documentclass[useAMS,usenatbib]{mn2e}
\usepackage{graphicx}
\usepackage{natbib}
\usepackage{amsmath,amssymb}
\usepackage{bbm}

\def\aj{AJ}%
\def\apj{ApJ}%
\def\apjs{ApJS}%
\def\aap{A\&A}%
\def\mnras{MNRAS}%
\def\bain{Bull.~Astron.~Inst.~Netherlands}%

\title[Stellar kinematics of M31]{Stellar kinematics using a third integral of motion: method and application on the Andromeda galaxy}

\author[R. Kipper et al]{R. Kipper$^{1,2}$\thanks{E-mail: rain.kipper@to.ee.},  P. Tenjes$^{1,2}$, O. Tihhonova$^{1,3}$, A. Tamm$^1$, E. Tempel$^{1}$\\
$^1$Tartu Observatory, Observatooriumi~1, 61602 T\~oravere, Estonia\\
$^2$Institute of Physics, University of Tartu, W. Ostwaldi 1, 51010 Tartu, Estonia\\
$^3$Laboratoire d'astrophysique, Ecole Polytechnique F{\'e}d{\'e}rale de Lausanne, Observatoire de Sauverny, CH-1290 Versoix, Switzerland}

\begin{document}

\date{}
\pagerange{\pageref{firstpage}--\pageref{lastpage}} \pubyear{2016}

\maketitle
\label{firstpage}

\begin{abstract}
We probe the feasibility of describing the structure of a multi-component axisymmetric galaxy with a dynamical model based on the Jeans equations while taking into account a third integral of motion. We demonstrate that using the third integral in the form derived by G. Kuzmin, it is possible to calculate the  stellar kinematics of a galaxy from the Jeans equations by integrating the equations along certain characteristic curves. In cases where the third integral of motion does not describe the system exactly, the derived kinematics would describe the galaxy only approximately.

We apply our method to the Andromeda galaxy, for which the mass distribution is relatively firmly known. We are able to reproduce the observed stellar kinematics of the galaxy rather well. The calculated model suggests that the velocity dispersion ratios $\sigma^2_z/\sigma^2_R$ of M31 decrease with increasing $R$. Moving away from the galactic plane, $\sigma^2_z/\sigma^2_R$ remains the same. The velocity dispersions $\sigma^2_\theta$ and $\sigma^2_R$ are roughly equal in the galactic plane.

\end{abstract}

\begin{keywords}
methods: analytical -- galaxies: kinematics and dynamics -- galaxies: individual: M31.
\end{keywords}

\section{Introduction}
\label{sec:introduction}

To determine the mass distribution of a galaxy, the Jeans equations \citep{Jeans:1915, Jeans:1922} are often used. These equations tie the density distribution and the gravitational potential of a stellar system to its kinematical characteristics: rotation velocities and velocity dispersions. A recent overview about different dynamical methods, including the ones based on the Jeans equations, is given by \citet{Courteau:2014}. For solving the Jeans equations, certain simplifying assumptions are usually made: stationarity and a certain symmetry. 

For spherically symmetric systems, good starting points are provided in the literature, e.g. for one integral model by \citet{Tremaine:1994}, for two integral models the Osipkov-Merritt model\footnote{Developed by \citet{Osipkov:1979} and \citet{Merritt:1985}.} \citep{Carollo:1995, Baes:2007}. For the general mass density distribution the Jeans equations were solved e.g. by \citet{Binney:1982}. 
It is worth to note that in some cases the Jeans equations can even be solved without assuming stationarity, enabling to study the evolution of galaxies or galaxy clusters \citep{Falco:2013}.

A typical assumption made for axisymmetric systems is that the phase density of stars is a function of two classical integrals of motion: the energy  and the angular momentum. For Jeans equations this means that the velocity dispersion distribution in a meridional $(R,z)$ plane of a galaxy is isotropic and aligned with the cylindrical coordinates\footnote{We designate the cylindrical coordinates in the usual form $(R, \theta ,z)$ with $z$ as the symmetry axis.} \citep[see e.g.][]{Binney:2008}. In the two Jeans equations the unknown functions are density, rotational velocity and the two velocity dispersion components (the Poisson equation relates the mass density and the gravitational potential). In principle, since mass density can be derived by deprojecting observed surface brightness distribution and rotation velocities and line-of-sight velocity dispersions can also be derived from observations, the mass distribution of a galaxy can be calculated \citep[see e.g.][]{Cinzano:1994, Cappellari:2006, Cappellari:2008, Williams:2009, Kipper:2012, Adams:2014}. In some special cases, the Jeans equations can even be solved analytically \citep{Smet:2014}. 

However, galaxy models based on the two integrals of motion do not always enable a satisfactory fit to the observed kinematics \citep{Binney:1990, vdMarel:1990, Merrifield:1991, Bottema:1993}. Moreover, the Hipparcos satellite \citep{1997ESASP1200.....E, 2007A&A...474..653V} observations of stellar velocities indicate that in the Solar neighbourhood, the calculated velocity dispersions along the three coordinate axes are not equal \citep{Dehnen:1998}. A similar conclusion was reached by \citet{Smith:2012} on the basis of the Stripe 82 data from the Sloan Digital Sky Survey. Away from the Milky Way plane the observed velocity dispersion ellipsoid is tilted towards the plane \citep{Siebert:2008, Binney:2014, Budenbernder:2014}. These observational facts can be explained by assuming that the phase density depends also on an additional, third integral of motion.

Besides, an elegant and powerful method for deriving mass distributions of galaxies has been developed by using the Schwarzschild orbit-superposition method \citep{Cretton:1999, Gebhardt:2003, Copin:2004, Valluri:2004, Krajnovic:2005, Cappellari:2006, Cappellari:2007, Thomas:2004, Thomas:2007, vandeVen:2008}. This method is independent from solving the Jeans equations. It is clear from these models (see also \citet{Ollongren:1962} for much earlier orbit calculations) that the third integral of motion has to be an essential part of the model construction.

If the third integral is taken into account, two additional unknown functions will appear in the Jeans equations, the third component of the velocity dispersions and the tilt angle of the velocity dispersion ellipsoid (see Section~\ref{sec:jeans_eq}). Hence, assumptions about the specific form of the third integral have to be made to solve the Jeans equations. For example, it is known that an analytical third integral form exists for the St\"ackel potential \citep[see e.g.][]{Binney:2008}. This was used by \citet{Batsleer:1994} and \citet{Famaey:2003} who constructed a Galaxy model as a sum of such potentials. A new, axisymmetric isochrone potential family with three integrals of motion was recently developed by \citet{Binney:2014b} and kinematical characteristics of the model were calculated. An actual galaxy can be modelled as a sum of these models. 

But even the usage of specific forms of the potential and the integral does not guarantee a satisfactory agreement between the model and the observations. For particular galaxies, the third integral may merely be a quasi-integral.

In the present paper we study stationary axisymmetric models in the framework of the Jeans equations and a third integral of motion. For the sake of the flexibility of the model, we do not want to limit ourselves with a specific density distribution form. Besides, we demand the form of the third integral to be applicable throughout the galaxy, not just in some special cases (e.g. close to the galactic plane or for nearly circular orbits). For these reasons we use the third integral in the analytical form derived by \citet{Kuzmin:1953,Kuzmin:1956} (see Section~\ref{sec:third_integral}) and develop a method to find an approximate solution that satisfies both Jeans equations. A galaxy is assumed to be a superposition of several components with density distribution derived from the (observed) surface brightness distribution. Using the Jeans equations, we calculate the distributions of stellar rotational velocity and velocity dispersion. The approach is similar to \citet{Tempel:2006}, where the third integral theory was used to model the Sombrero galaxy. The present work is the extension of the previous work by forfeiting a relation that holds only near to the plane of the galaxy. 

Our method resembles the method recently developed by \citet{Bien:2015} and used for modelling the disc component of the Milky Way embedded in a dark matter halo. In the referred paper, a form for the third integral was selected and applied along stellar orbits, determining the best-fit integral value for each orbit. It was found that the used integral form suited to the orbits rather well (i.e. remained nearly constant along the orbits). In contrast, we seek a fixed but simple form of the integral providing the best fit to the observed kinematics for the entire galaxy.

Since the Andromeda galaxy is nearby and has been observed countless times, it is a popular test body for galaxy models. In recent years \citet{Geehan:2006}, \citet{Seigar:2008}, \citet{Chemin:2009}, \citet{Corbelli:2010}, \citet{Tamm:2012} have tested a variety of kinematical modelling techniques on it, gaining mostly consistent results, which allows us to consider the general density distribution of M31 to be sufficiently settled. For test purposes we selected a simple version of bulge + disc + dark matter halo model developed in \citet{Tamm:2012}. 

The paper is organised as follows. In Section~\ref{sec:method} we present the Jeans equations in a form suitable for the present study, introduce the third integral of motion and provide a recipe for solving the Jeans equations. In Section~\ref{sec:m31} we apply our model to the Andromeda galaxy and compare the calculated rotation velocities and velocity dispersions with observations. A discussion of the results and our conclusions are given in Section~\ref{sec:conclusions}.

\section{Method}
\label{sec:method}

\subsection{Jeans equations}
\label{sec:jeans_eq}

In the most general form the Jeans equations can be written as \citep[see ][]{Binney:2008}
\begin{equation}
\rho\frac{\partial \overline{v}_j}{\partial t}+\rho\overline{v}_i\frac{\partial\overline{v}_j}{\partial x_i}=-\rho\frac{\partial\Phi}{\partial x_j}-\frac{\partial(\rho\sigma^2_{ij})}{\partial x_i},
\end{equation}
where $x_i$ and $v_i$ represent Cartesian coordinates and velocities, $t$ is time, $\rho$ and $\Phi$ denote the mass density and the gravitational potential, and $\sigma_{ij}^2$ are the components of the velocity dispersion tensor
\begin{equation}
\sigma_{ij}^2=\overline{(v_i-\overline{v_i})(v_j-\overline{v_j})}=\overline{v_iv_j}-\overline{v_i}~\overline{v_j}.
\end{equation} 
Density $\rho$ and gravitational potential $\Phi$ are tied through the Poisson's equation $\nabla^2\Phi=4\pi G\rho$, where $G$ is the gravitational constant. 

As our aim is to develop a model for a multi-component galaxy, all the kinematic variables and densities will be considered per component, except for the gravitational potential, which contains the contribution of all the components. 

We assume that mass distribution in a galaxy can be approximated with an axially symmetric model. In this case the two mixed components of the velocity dispersion tensor are zero
\begin{equation}
	\sigma^2_{R\theta}=\sigma^2_{z\theta}=0. \label{eq:sega12}
\end{equation}

If the phase density of a stellar system is a function of the two classical integrals of motion, the energy and the angular momentum integrals, the third mixed component $\sigma^2_{Rz}$ of the dispersion tensor will also be zero and the velocity ellipsoids (the nonzero diagonal components of the tensor) will be aligned with the three cylindrical coordinate axes. However, if the phase density is also a function of a third integral of motion, in addition to the two classical ones, the mixed dispersion tensor component $\sigma^2_{Rz}$ will be nonzero. In this case the velocity dispersion tensor takes the diagonal form in some other coordinates. One axis of the velocity ellipsoid would still coincide with the $\theta$ axis of the cylindrical coordinates (the ellipsoid lies in a meridional plane of the galaxy), but in the $R$-$z$ plane, the velocity ellipsoid would be tilted by an angle $\alpha$ with respect to the galactic plane or the $R$ axis (see Fig.~\ref{fig:alpha} for illustration; do not pay attention to the elliptical coordinate set at this point). It can be shown that $\sigma^2_{Rz}$ is related to the tilt angle $\alpha$ via the relation 
\begin{equation}
	\sigma^2_{Rz}=\gamma(\sigma_{RR}^2-\sigma_{zz}^2), 
	\quad \mathrm{where} \quad \gamma=\frac{1}{2}\tan{2\alpha}.
	 \label{eq:sega3}
\end{equation}
Denoting $\sigma^2_z\equiv\sigma^2_{zz}$, $\sigma^2_R\equiv\sigma^2_{RR}$, and $\sigma^2_{\theta}\equiv\sigma^2_{\theta\theta}$ for brevity, the shape of the velocity ellipsoid can be described by the axial ratios of the ellipsoid
\begin{eqnarray}
	\label{eq:kz.def}
	k_z&\equiv&\sigma_z^2/\sigma_R^2,\\
	\label{eq:kt.def}
	k_\theta&\equiv&\sigma_\theta^2/\sigma_R^2.
\end{eqnarray} 
 
Assuming a stationary axisymmetric mass distribution and using relations (\ref{eq:sega12}), (\ref{eq:sega3}) and designations (\ref{eq:kz.def}), (\ref{eq:kt.def}), the Jeans equations can be written in cylindrical coordinates 
\begin{eqnarray}\label{Jeanseq1}
	\frac{\partial(\rho\sigma_R^2)}{\partial R} + \left( \frac{1-k_\theta}{R}+\frac{\partial\kappa}{\partial z} \right)\rho\sigma_R^2 + \kappa\frac{\partial(\rho\sigma_R^2)}{\partial z} =\\
	= -\rho\left( \frac{\partial\Phi}{\partial R}-\frac{V_\theta^2}{R} \right)\nonumber, \\ \label{Jeanseq2} 
	\frac{\partial(\rho\sigma_z^2)}{\partial z} + \left( \frac{\xi}{R}+\frac{\partial\xi}{\partial R} \right)\rho\sigma_z^2 + \xi\frac{\partial(\rho\sigma_z^2)}{\partial R} = -\rho\frac{\partial\Phi}{\partial z},
\end{eqnarray}
where
\begin{eqnarray}
	\label{eq:kappa.def}
	\kappa &\equiv& \gamma(1-k_z),\\
	\label{eq:xi.def}
	\xi &\equiv& \kappa/k_z.
\end{eqnarray} 
In cylindrical coordinates, one of the Jeans equations turns to identity. The Jeans equations given in the latter form are convenient for our further calculations.

At this point, we have unknown functions $\rho$, $\Phi$, $\sigma_R^2$, $\sigma_{\theta}^2$, $\sigma_z^2$, $\gamma$ and the rotational velocity $V_{\theta} \equiv \overline{v}_{\theta}$ each being a function of both $R$ and $z$, but only two Jeans equations and the Poisson's equation. This is not sufficient for solving the system. Below we show that an expression for the shape and the tilt angle of the velocity ellipsoid can be derived from the theory of a third integral of motion and the resulting system is solvable.

\subsection{Third integral of motion}
\label{sec:third_integral}

In the Solar neighbourhood, none of the three diagonal components of the velocity dispersion tensor are  equal \citep{Dehnen:1998}. This indicates that at least in the Solar neighbourhood, beside the two classical energy and angular momentum integrals
\begin{eqnarray}
	I_1 &=& v_R^2+ v_{\theta}^2 + v_z^2 - 2\Phi , \label{i1} \\
	I_2 &=& R v_{\theta},
	\label{i2}
\end{eqnarray}
a third integral of motion has to exist. Additionally, \citet{Valluri:2004}, \citet{Cappellari:2006} and \citet{Vasiliev:2015} have modelled galaxies with the Schwarzschild orbit superposition method and showed that in elliptical galaxies a third, nonclassical integral exists. Further arguments supporting the existence of a third integral were given in the introduction.

\begin{figure}
  	\includegraphics[width=84mm]{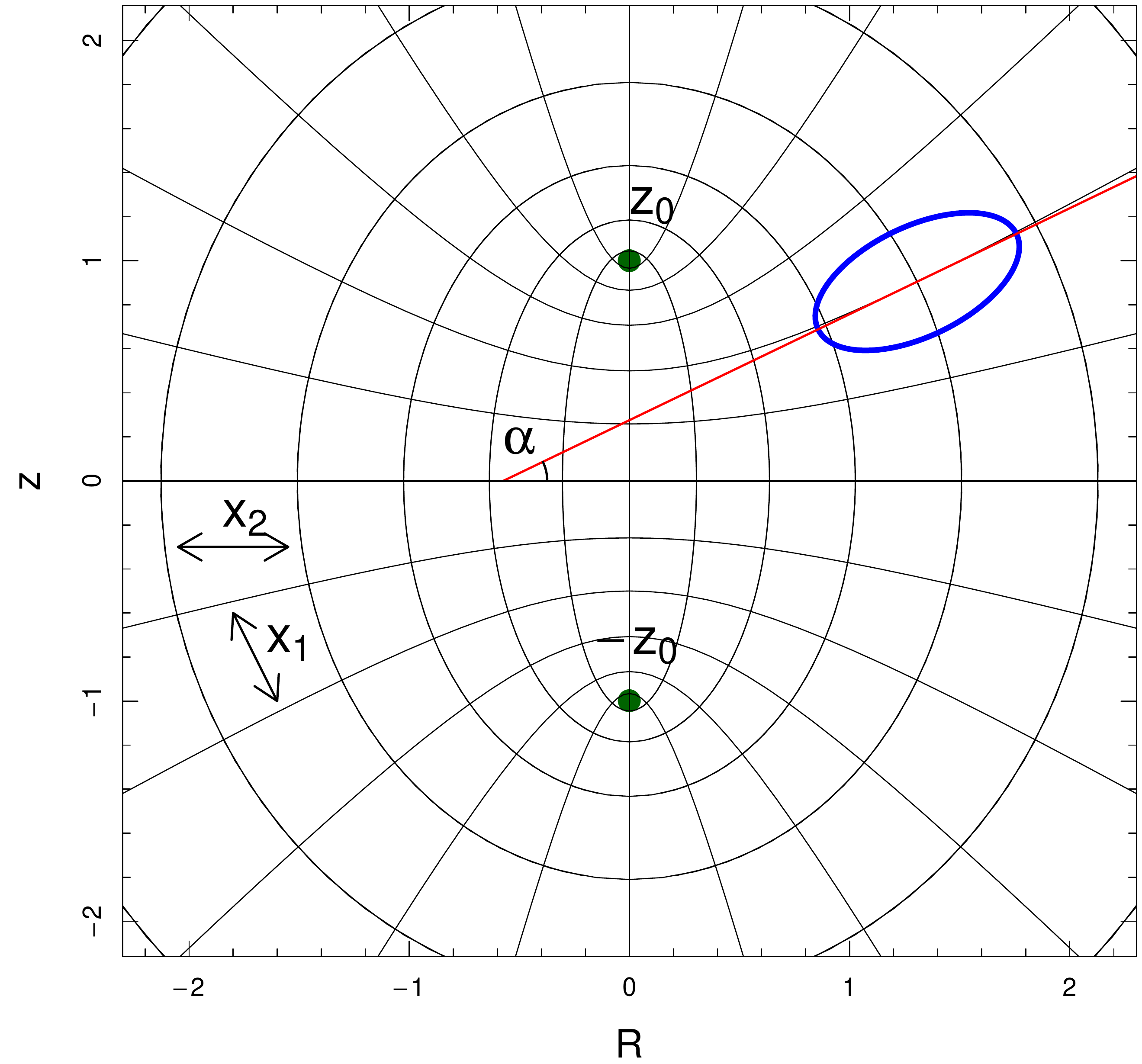}
 	\caption{Geometrical representation of elliptical coordinates ($x_1$ and $x_2$) and the tilt angle $\alpha$ of the dispersion tensor in a meridional plane of galaxy. The foci of coordinates are marked with $\pm z_0$. A velocity ellipsoid is shown with the thick blue line. The galaxy plane is parallel with $R$ coordinates and is located at $z=0$, galaxy centre is located at the origin of coordinates. }
 	\label{fig:alpha}
\end{figure}

We have chosen to use the third integral of motion in the form derived by \citet{Kuzmin:1953, Kuzmin:1956}. The aim of these papers was to gain the most general form for this integral. Kuzmin started by assuming that it is a quadric function with respect to the velocities -- otherwise too many restrictions for the potential would appear. His approach led to the following form for the third integral:
\begin{equation}
	I_3 = (Rv_z-zv_R)^2 + z^2v_\theta^2+z_0^2(v_z^2-2\Phi^*). \label{i3}
\end{equation} 
The function $\Phi^*$ is related to the gravitational potential via relations 
\begin{eqnarray}
	z_0^2\frac{\partial\Phi^*}{\partial R}&=&z^2\frac{\partial \Phi}{\partial R}-Rz\frac{\partial\Phi}{\partial z},\\
	z_0^2\frac{\partial\Phi^*}{\partial z}&=&(R^2+z_0^2)\frac{\partial\Phi}{\partial z}-Rz\frac{\partial\Phi}{\partial R}.
\end{eqnarray}
In this case the velocity dispersion tensor is in the diagonal form in elliptical coordinates defined as
\begin{eqnarray}
	x_1^2&=&\frac12[\Omega+\sqrt{\Omega^2-4z^2z_0^2}],\\
	x_2^2&=&\frac12[\Omega-\sqrt{\Omega^2-4z^2z_0^2}],
\end{eqnarray} 
where $\Omega=R^2+z_0^2+z^2$ and the parameters $\pm z_0$ correspond to the foci of coordinates $(x_1, x_2)$ (see Fig.~\ref{fig:alpha}). The inclination angle $\alpha$ between the ellipsoid and the plane of the galaxy is given as
\begin{equation}
	\gamma = \frac12\tan2\alpha = \frac{Rz}{R^2+z_0^2-z^2}
	\label{eq:gamma.def}
\end{equation}
i.e. is determined by the $z_0$ value. 

Now, the integrals of motion (\ref{i1}) -- (\ref{i3}) can be written as
\begin{eqnarray}
	I_1 &=& v_1^2+ v_{\theta}^2 + v_2^2 - 2\Phi , \label{i11}\\
	I_2 &=& R v_{\theta}, \label{i22}\\ 
	I_3 &=& x_1^2v_1^2+x_2^2v_2^2+x_1x_2v_{\theta}^2-2z_0^2\Phi^*.
	\label{i33} 
\end{eqnarray}
Following \citet{Einasto:1970, Einasto:1972}, and assuming that the velocity distribution remains ellipsoidal, the phase density must be quadratic with respect to the velocities. Thus the phase density must be a function of the integrals in the form of a combination
\begin{equation}
	Q = a_1 I_1 + a_2 I_3 + 2\frac{b_1}{z_0} I_2 + \frac{b_2}{z_0^2} I_2^2,
\end{equation}
where $I_1$, $I_2$, $I_3$ are the integrals of motion (\ref{i11}) -- (\ref{i33}) and $a_1$, $a_2$, $b_1$ and $b_2$ are constants. As the integrals of motions can be multiplied by arbitrary constants, we multiply all of them with $a_1^{-1}$, effectively taking $a_1 = 1$, without loss of generality. This yields for the velocity dispersion ratios
\begin{eqnarray}
	k^*_{12}&=&\frac{z_0^2+a_2x_2^2}{z_0^2+a_2x_1^2},	\label{eq:k12}\\
	k^*_{13}&=&\frac{z_0^2+a_2x_2^2}{z_0^2+a_2z^2+b_2R^2}, \label{eq:k13}
\end{eqnarray}
or the dispersion ratios in the cylindrical coordinate axes direction
\begin{eqnarray}
	k^*_z&=&\frac{\sin^2\alpha+k_{12}\cos^2\alpha}{\cos^2\alpha+k_{12}\sin^2\alpha}=\frac{\tan^2\alpha+k_{12}}{1+k_{12}\tan^2\alpha},	\label{eq:kz}\\
	k^*_\theta&=&\frac{k_{13}}{\cos^2\alpha+k_{12}\sin^2\alpha}=\frac{k_{13}(1+\tan^2\alpha)}{1+k_{12}\tan^2\alpha}. 	\label{eq:kt}
\end{eqnarray} 
The asterisk $^*$ denotes that the corresponding expressions are derived from the Kuzmin's third integral of motion. 

Now, instead of the four unknown functions $\sigma_R^2$, $\sigma_{\theta}^2$, $\sigma_z^2$ and $\gamma$ describing the velocity dispersion tensor, we have one unknown function, e.g. $\sigma_R^2$, and three unknown constants $z_0$, $a_2$ and $b_2$. Note that, assuming the third integral theory, the velocity ellipsoid is oriented along elliptical coordinates given with foci $z_0$ at every point of the galaxy. In general, if $z_0$ is a function of $R$ and $z$ (i.e. the tilt of the velocity ellipsoid is not tied to fixed elliptical coordinates) instead of being a constant, the third integral theory is only an approximation and the third integral is actually a quasi-integral.

\subsection{Solving the Jeans equations}
\label{subsec:vabad.parameetrid} 

One way to solve the Jeans equations in the case of triaxial velocity distribution was proposed by \citet{Kuzmin:1987}. Kuzmin studied the third Jeans equation in ellipsoidal coordinates. By choosing a suitable form for the dispersion ratio $\sigma_1 / \sigma_2$ he derived an expression for $\sigma_1$ or $\sigma_2$ where integrations were made along hyperbolae expressed analytically in a simple form. The first Jeans equation can be solved in a similar way.

\citet{Evans:1991} solved the Jeans equations for St\"ackel-type potential. In this case the dispersion ratio $\sigma_1/\sigma_2$ has a simple form and the Jeans equations can be represented as a system of four simple differential equations with partial derivatives. This system can also be solved with a method of characteristics but integration lines must be calculated numerically.

In the present paper we use a similar method. The two Jeans equations form a system as both equations contain variable $k_z$ and derivatives with respect to $R$ and $z$. However, by using the Kuzmin third integral form and an assumption that velocity distribution remains ellipsoidal, we may specify expressions for the velocity ellipsoid parameters $k_z$, $k_\theta$ and $\alpha$ (see Eqs.~\eqref{eq:gamma.def}, \eqref{eq:kz}  and  \eqref{eq:kt}). This allows us to solve the Jeans equations (\ref{Jeanseq1}) and (\ref{Jeanseq2}) with the method of characteristics (see Appendix \ref{sec:the_solving} for details). 

Resulting expressions for dispersions $\sigma^2_R$ and $\sigma^2_z$ are
\begin{eqnarray}
	\label{eq:solve_jeans_sigma_r}
	\sigma^2_R(R,z)\!\!\!&=&\!\!\!\frac{1}{\rho}\int\limits_R^\infty\rho(r,z)(1-\beta^2)\frac{\partial\Phi(r,z)}{\partial r} e^{\int\limits^r_Rp(r^*,z)\mathrm{d}r^*} \mathrm{d}r , \\
	\label{eq:solve_jeans_sigma_z}
	\sigma_z^2(R,z)\!\!\!&=&\!\!\!\frac{1}{\rho} \int\limits^\infty_z \rho(R,z') \frac{\partial\Phi(R,z')}{\partial z'} e^{\int\limits_z^{z'}g(R,z^*)\mathrm{d}z^*} \mathrm{d}z' ,
\end{eqnarray} 
where the functions $p$ and $g$ are denoted as 
\begin{eqnarray}
	p&=&\frac{1-k^*_\theta}R+\frac{\partial\kappa^*}{\partial z},\\
	g&=&\frac{\xi^*}R+\frac{\partial\xi^*}{\partial R} 
\end{eqnarray}
and integration goes along the characteristic curves. Again, the asterisks (also in $\xi$ and $\kappa$) denote that the variables are calculated based on the Kuzmin third integral approximation. We introduce also a function $\beta$ defined as
\begin{equation}
	\beta^2 = \frac{V_\theta^2}{v_c^2} =V_\theta^2  \left(R \frac{\,\partial\Phi}{\partial R} \right)^{-1}.
\end{equation}
Function $\beta (R,z)$ can be constrained by observations (e.g. using data from the Gaia mission in the near future, by comparing the rotation of a test population of objects with rotation velocities of the thin gas disc).

In principle, if we know the free parameters ($a_2$, $b_2$ and $z_0$), we can use either one of the Jeans equations to calculate all the kinematical variables (see below). The overall result must not depend on which Jeans equation to use. Unfortunately, we do not know the free parameters and need a way to constrain them. We use the before mentioned condition, that for a correct set of free parameters, the Jeans equations must yield concordant velocity dispersions: we calculate the radial velocity dispersion ($\sigma^2_R$) from the first Jeans equation and the vertical velocity dispersion ($\sigma_z^2$) from the second one. Using the shape of the velocity ellipsoid ($k^*_z$) that was used in the Jeans equations, we can check whether the found dispersions are consistent
\begin{equation}
	\sigma_R^2 = \sigma_z^2/k^*_z.\label{eq:consistency}
\end{equation}
This equation (or an equivalent equation $k_z=k^*_z$) turns into identity only if the Jeans equations and the third integral of motion are simultaneously satisfied. This gives a way to find the free parameters in the model. It is implemented by constructing a cost function 
\begin{equation}
	\label{eq:chisq}
	\chi^2=\int\limits_0^\infty\int\limits_0^\infty\left[\sigma_{R}(R,z)-\sigma_{z}(R,z)/\sqrt{k^*_z}\right]^2\mathrm{d}R\mathrm{d}z
\end{equation}
and minimising it.  Alternatively, more elaborate matching techniques can be used. If $\chi^2=0$, then the solution is exact, otherwise it is only an approximation.

We would like to stress that the Kuzmin third integral form is not the only possible one and it does not have to be necessarily exact for a galaxy with complicated overall density distribution. Thus it is possible that the Jeans equations and the third integral of motion will not match perfectly\footnote{If the solution to the Jeans equations is an approximation (i.e. not exact) the system is likely in a quasi-equilibrium state.}. For practical exercises, one can still use the method, but must bear in mind that to some extent, the results would slightly depend on whether one calculates kinematics based on $\sigma^2_R$ or $\sigma^2_z$ (see Fig.~\ref{fig:rot.curve}).

Once the value for either $\sigma^2_R$ or $\sigma^2_z$ has been found, the other components can be calculated using the third integral approximation. The formulae for other components of $\sigma^2_R$ based kinematics are $\sigma^2_z = k_z^*\sigma^2_R$, $\sigma^2_\theta = k_\theta^*\sigma^2_R$ and in the case of $\sigma^2_z$ based kinematics $\sigma^2_R = \sigma^2_z/k_z^*$, $\sigma^2_\theta = \sigma^2_zk_\theta^*/k_z^*$. The tilt of the ellipsoid comes from Eq.~(\ref{eq:gamma.def}). 
The non-tilted velocity ellipsoid parameters can be found with the following equations:
\begin{eqnarray}
	\sigma_1^2 &=& \frac{\sigma^2_z\sin^2\alpha - \sigma_R^2\cos^2\alpha}{\sin^2\alpha-\cos^2\alpha},\\
	\sigma_2^2 &=& \frac{\sigma^2_R\sin^2\alpha - \sigma_z^2\cos^2\alpha}{\sin^2\alpha-\cos^2\alpha}.
\end{eqnarray}
In the case of $\alpha=45^\circ$, there is a 0/0 indetermination and the non-tilted ellipsoid parameters can be found using the dispersion ratio $k_{12}$.

\subsection{Line of sight velocity distribution}\label{sec:losvd}

Observationally, we measure the line-of-sight velocity component of stellar velocities. Ignoring the effects of light absorption and scattering by the interstellar dust (a good approximation for older stellar systems), the stellar populations of galaxies can be considered transparent, thus the spectral lines contain the contribution of all stars along a given line of sight. Therefore, to be comparable with observational data of actual galaxies, we need to project the modelled velocity distribution (both $V_\theta$ and $\sigma^2$) to the viewing direction and integrate along the line of sight through the entire galaxy.

Our model gives velocity dispersions along the cylindrical coordinate axes, thus a simple coordinate rotation is needed to compute the dispersion in the line of sight direction. Let $X$ and $Y$ be distances along the major and minor axes of the plane-of-the-sky projection of the galaxy, respectively, and $\delta$ denote the inclination angle, defined as the angle between the rotation axis and the line of sight ($90^\circ$ corresponds to an edge-on galaxy). In order to find the velocity dispersion along a sightline, the most simple way is to find the non-tilted dispersion tensor shape and project the value of the velocity ellipsoid to the line of sight. Once the ellipsoidal coordinate-aligned ellipsoid is found, one can get the line-of-sight projection $(\sigma^2_\mathrm{los})$ using:
\begin{eqnarray}
	\sigma^2_\mathrm{los} &=& \sigma^2_\mathrm{mer}\frac{R^2-X^2\sin^2\delta}{R^2}+\sigma^2_\theta\frac{X^2\sin^2\delta}{R^2},\\
	\sigma^2_\mathrm{mer} &=& \sigma_1^2\cos^2\zeta + \sigma_2^2\sin^2\zeta,\\
	\zeta &=&  \alpha - \arctan\left(\frac{R}{\tan{\delta}\sqrt{R^2-X^2}}\right),
\end{eqnarray}
where $\zeta$ is the angle in the meridional plane toward the projected observer direction, which is a combination of the tilt of the velocity ellipsoid and the angle between the line of sight and the cylindrical coordinate set. 
Similarly, we need to extract the line-of-sight component from the rotational velocity $V_\theta$:
\begin{equation}
	V_\mathrm{los}(R,z)=V_\theta\frac XR\sin\delta.
\end{equation}

The observed spectral line shape at a given point in the plane of the sky ($X,Y$) effectively forms as the luminosity-weighted sum of the velocity distributions of each location along the corresponding line of sight. This velocity distribution $I(v)$ along a given line of sight can thus be calculated as:
\begin{eqnarray}
I(X,Y,v)=\int\limits_{-\infty}^\infty N\{V_\mathrm{los}[R'(z), z], \sigma_\mathrm{los}[R'(z), z]\} \cdot\\\cdot \frac{\rho[R'(z), z]}{\Upsilon}\frac{1}{\cos{\delta}}\mathrm{d}z,\nonumber
\end{eqnarray}
where 
\begin{equation}
	R'(z)=\sqrt{X^2+\left(z\tan{\delta}-\frac{Y}{\cos{\delta}}\right)^2}
\end{equation}
is a function linking $z$ and $R$ coordinates at each point along the sightline, $N$ denotes the normal distribution, and $\Upsilon$ the mass-to-light ratio of the galaxy component. 

For galaxy models with multiple components (a bulge, a disc, etc.), $I$ must be summed over components
\begin{equation}
	I_\mathrm{combined}(X,Y,v) = \sum\limits_i I_i(X,Y,v).
\end{equation}
To compare the resulting line-of-sight velocity distributions with observations, one needs to approximate the $I_\mathrm{combined}$ values with normal or Gauss-Hermite profiles.

\section{Applying the method to M31}
\label{sec:m31}

\subsection{Density profile}

To verify the applicability of the model described in the previous section, we apply the model on the well-studied nearby galaxy M31. In \citet{Tamm:2012}, the mass distribution of M31 is approximated with a usual three-component model: stellar bulge + stellar disc + dark matter halo\footnote{In \citet{Tamm:2012} also a more sophisticated (five stellar components) mass distribution model is derived. In the present analysis we limit ourselves with a simplified bulge\,+\,disc model for the stellar components, mostly to keep the number of free parameters minimal during calculations. Besides, the bulge and the disc dominate the stellar mass budget of the M31; additional components would have a negligible effect on the mass distribution and gravitational potential.}. The parameters of the stellar components in the latter model were found by fitting dust-corrected surface brightness distributions \citep[derived in][]{2010A&A...509A..91T, 2011A&A...526A.155T} and assuming a constant mass-to-light ratio for each component. Dark matter density distribution is estimated by subtracting the stellar mass contribution from the observed gas rotation curve and the enclosed mass estimates in the outer regions. It is important to notice that for the referred mass distribution model construction, no stellar kinematics (neither rotation velocity nor velocity dispersion) information is used. As a consequence, the mass distribution derivation is independent of the observed stellar kinematics. 

\begin{table}
	\centering
 	\caption{Mass distribution model parameters of M31, taken from \citet{Tamm:2012}. The parameters for each component correspond to the density distribution Eq.~(\ref{eq:einasto}).}
	\label{comp.par.table}
	\begin{tabular}{lccccc}
		\hline
		Component & $a_c$ & $q$ & $N$ & $\rho_c$ & $M$ \\
		          & kpc   &     &    & $M_\odot\mathrm{pc}^{-3}$ & $10^{10}M_\odot$ \\
	    \hline
		\hline
	    Bulge & 2.025 & 0.73 & 4.0 & 0.220  & 4.9 \\
		Disc  & 11.35 & 0.10 & 1.0 & 0.017  & 4.8 \\
		Dark matter & 178.0 & 1.00 & 6.0 & $8.12\times10^{-6}$ & 205.6 \\
		\hline
	\end{tabular}
\end{table}

In \citet{Tamm:2012} the density distribution of M31 is given as a superposition of the Einasto profiles \citep{1969Afz.....5..137E}
\begin{equation}
	\label{eq:einasto}
	\rho(a)=\rho_c \exp{\left\{-d_N\left[ \left( \frac{a}{a_c} \right)^{1/N}-1\right]\right\}},
\end{equation}
where $\rho$ is density, $N$ is the Einasto index, which sets the shape of the distribution (similar to the S\'ersic index), $d_N$ is a function of $N$, $a=\sqrt{R^2+z^2/q^2}$ is the equivalent of distance in a spherically symmetric model, and $\rho_c=hM\exp(-d_N)k^3d_N^{3N}/(4\pi qa_c^3)$ defines density at $a_c$, where $M$ and $q$ are the mass and the flatness of the component, respectively, and $h$ and $k$ are normalising constants; Appendix~B of \citet{Tamm:2012} gives the definition of the normalising constants (and the relations between various popular forms of the Einasto's distribution). The values of the parameters of each M31 component (as used below) are given in Table~\ref{comp.par.table}.

To solve the Jeans equations, we also need to calculate derivatives of the  gravitational potential. For the Einasto's profile the derivatives can be expressed as \citep[see][]{Tenjes:2001}
\begin{eqnarray}
	\label{eq:pot_r}
	\frac{\partial\Phi}{\partial R}=R\frac{GhMk^3d_N^{3N}}{(ea_c)^3}\int\limits_0^{\arcsin e}\rho^*(a)\sin^2(x)\,\mathrm{d}x,\\
	\label{eq:pot_z}
	\frac{\partial\Phi}{\partial z}=z\frac{GhMk^3d_N^{3N}}{(ea_c)^3}\int\limits_0^{\arcsin e}\rho^*(a)\tan^2(x)\,\mathrm{d}x,
\end{eqnarray}
where $e=\sqrt{1-q^2}$ is eccentricity, $\rho^*=\exp\left[ -d_N\left( \frac{a}{a_c} \right)^{1/N} \right]$ and $a^2=\frac{\sin^2(x)}{e^2}\left( R^2+\frac{z^2}{\cos^2(x)} \right)$. Eqs.~(\ref{eq:pot_r}) and~(\ref{eq:pot_z}) can be used if $q<1$. For spherical systems (e.g. the dark matter halo for M31), the derivatives of the gravitational potential are
\begin{eqnarray}
	\frac{\partial\Phi(R,z)}{\partial R} &=& R\frac{GhMk^3d_N^{3N}}{a_c^3}\int\limits^1_0\rho^*(a)x^2\textrm{d}x,\\
	\frac{\partial\Phi(R,z)}{\partial z} &=& z\frac{GhMk^3d_N^{3N}}{a_c^3}\int\limits^1_0\rho^*(a)x^2\textrm{d}x,
\end{eqnarray}
where $a^2=x^2(R^2+z^2)$.

\subsection{Solving of the Jeans equations for M 31}
\label{sec:m31_jeans}

To apply our kinematical model to M31 we have to find the free parameters of the model. The velocity dispersions were calculated (i.e. Jeans equations were solved) from Eqs. (\ref{eq:solve_jeans_sigma_r}) and (\ref{eq:solve_jeans_sigma_z}). This was done separately for bulge and disc. To simplify calculations we approximated the function $\beta$ as a superposition of two constant values, one for the bulge and one for the disc:
\begin{equation}
	\beta^2 = V_\theta^2 / v_c^2 = \mathrm{const.}
\end{equation}
Note that effectively, $\beta$ still remains a function of $R$ and $z$ for the galaxy as a whole, since the  contribution of either component varies with mass density.

\begin{figure}
  	\includegraphics[width=84mm]{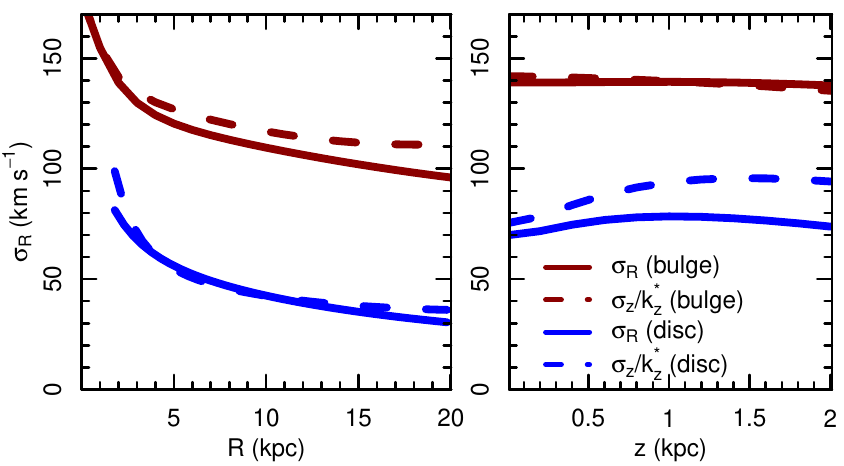}
 	\caption{Velocity dispersions ($\sigma_R$) along a line slightly off the major axis (left panel) and the minor axis (right panel) of M31 for two cases. Solid lines show the dispersions calculated directly from the first Jeans equation Eq.~(\ref{eq:solve_jeans_sigma_r}), dashed lines show the dispersions found using the second Jeans equation Eq.~(\ref{eq:solve_jeans_sigma_z}). Red colour corresponds to the bulge, blue colour to the disc component. Note that in the case of the bulge, the density drops rapidly and the apparent discrepancy seen on the left panel has negligible impact on actual calculations. }
 	\label{fig:disp.match}
\end{figure} 

The free parameters to be determined for the bulge and the disc thus also include $\beta$. These parameters are found by demanding that solutions of the two Jeans equations are mutually consistent and also consistent with the observed stellar rotation velocities. For this we minimise Eq.~(\ref{eq:chisq}) by sampling through the free parameter space ($a_2$, $b_2$, $\beta$ and $z_0$) to find the parameter set that gives the smallest value to $\chi^2$. During the disc $\chi^2$ calculations, we excluded regions where the bulge dominated the density to improve the overall quality of the model (this has very little effect when comparing with observations). For fitting the free parameters we use the Bayesian analysis tool \emph{multinest} \citep{MN2,MN1,MN3}, which finds the most probable set of parameters and also their posterior widths (statistical uncertainties). We use wide and uniform priors for all the parameters. The resulting bulge and disc parameters, that make the Jeans equations consistent with the third integral, are given in Table~\ref{z0table} together with uncertainties. One should notice that the high value of $z_0$ changes the velocity ellipsoid toward isotropic shape (see Eqs.~(\ref{eq:k12}, \ref{eq:k13})), reducing the importance of $a_2$ and $b_2$. 

In general, the form of Eq.~\eqref{eq:k12} indicates, that there could be a degeneracy between the parameters $z_0$ and $a_2$. Posterior distribution of the parameters confirmed it to exist, but not high enough to influence the results of the model. The parameter uncertainties given in Table 2 do not take into account the degeneracies between the fitted parameters, they correspond to the best fitted model\footnote{During our model calculations, we tried different approaches that lead to slightly different parameter values. Hence, the parameter values cannot be determined uniquely. However, the observational quantities calculated from the model were not affected by the different parameter values, hence the model is robust with respect to the observational quantities.}.

\begin{figure*}[h]
 	\includegraphics[width=176mm]{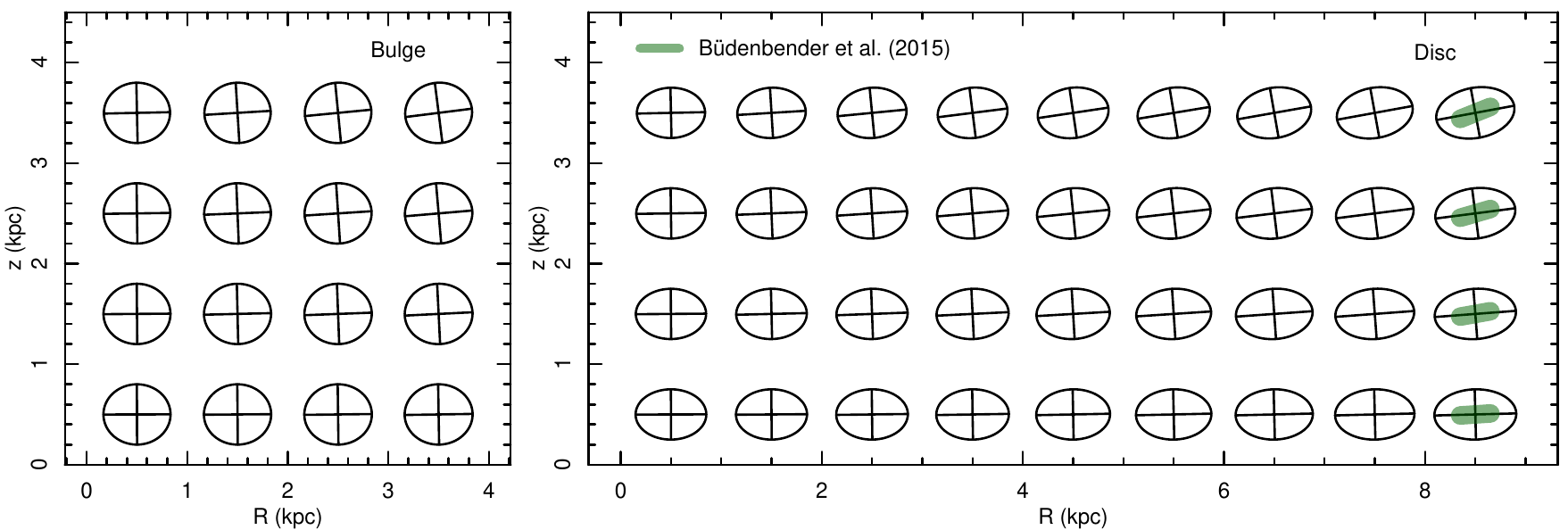} 
 	\caption{Calculated orientations and axial ratios of the velocity ellipsoids of the bulge (left panel) and the disc (right panel) components of M31 in a meridional plane of the galaxy. For a comparison, thick green dashes  give tilt angles of the velocity ellipsoids in our Galaxy \citep{Budenbernder:2014}.}
 		\label{fig:orient}
\end{figure*} 

\begin{table}
	 \centering
	 \caption{Best-fit values of the free parameters of the model. Parameters $a_2$, $b_2$ and $z_0$ describe the shape and orientation of the velocity ellipsoid, $\beta$ sets rotation velocities of the stellar components. The indicated errors are standard deviations derived directly from the Bayesian analysis tool \emph{multinest}.}
	 \label{z0table}
 	\begin{tabular}{l|cc}
		\hline
		Parameter & Bulge & Disc\\
		\hline
		\hline
		$a_2$    & $0.11 \pm 0.004$ & $0.41 \pm 0.002$ \\
		$b_2$    & $0.01 \pm 0.04$ & $0.04 \pm 0.02$ \\
		$z_0$ & $9.8 \pm 4.0$ & $11.5 \pm 0.3$ \\
		$\beta$  & $0.32 \pm 0.03 $ & $0.95 \pm 0.04 $ \\
		\hline
 	\end{tabular}
\end{table} 

Fig.~\ref{fig:disp.match} shows radial velocity dispersions (both $\sigma^2_R$ and $\sigma_z^2/k^*_z$) along the major axis and along the minor axis of M31, derived from either of the Jeans equations. A good match between the corresponding distributions would mean that the model assumptions work well and the third integral holds. A not so good match would mean that either the Kuzmin third integral can only be an approximation, (i.e. it is a quasi-integral), or hint that the system is only close to equilibrium\footnote{\citet{Kirk:2015} found that the centre of the gas ring is offset compared to the galactic centre, indicating a possible perturbed state.}. In either interpretation, there is a question whether the solution describes the real galaxy, and results must be used with  precaution. In the current case, the dispersions $\sigma^2_R$ and $\sigma_z^2/k^*_z$ in the bulge component are in a good agreement. For the disc component, notable differences occur because the velocity ellipsoid is flatter and the influence (and errors) of the ellipsoid orientation increases. However, integration along the line of sight suppresses these deviations significantly (black and grey lines in Fig.~\ref{fig:rot.curve}). As these lines are very close, we conclude, that the approximation is adequate.

In general, match between $\sigma^2_R$ and $\sigma_z^2/k^*_z$ can be improved assuming that the third integral of motion is a quasi-integral and let the foci  $z_0$ of the elliptical coordinates to be a weak function of $R$ and $z$. In present paper we decided to keep $z_0$ constant and found that in the case of M31 this is a satisfactory or even a good approximation (see Section~\ref{sec:m31_obs}).

Fig.~\ref{fig:orient} illustrates the shape and orientation of the modelled dispersion tensor along a meridional plane of the galaxy. The dispersion ellipsoid stays almost spherical throughout the bulge component. In the disc, the ellipsoid is close to spherical only near the rotation axis and  flattens out towards the edge, whereas the tilt angle of the ellipsoid increases with distance from the disc plane. 

\begin{figure}
  	\includegraphics[width=84mm]{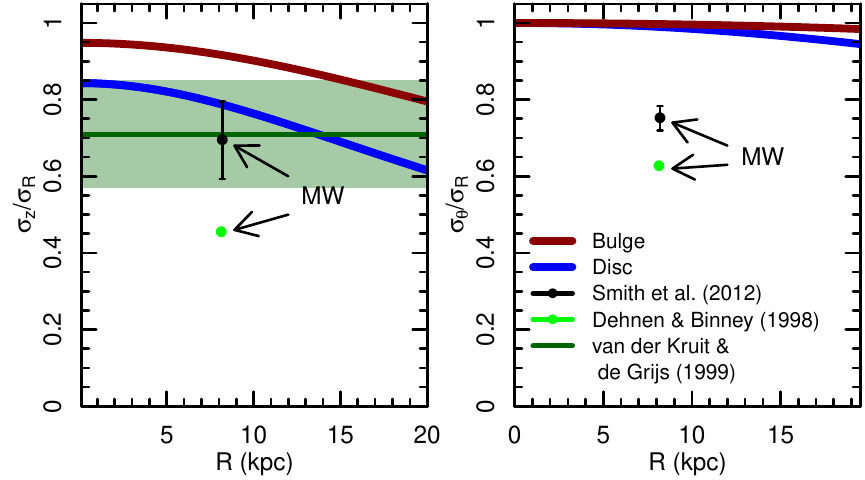}
  	\caption{Velocity dispersions ratios $\sigma_z/\sigma_R$ (left panel) and $\sigma_\theta/\sigma_R$ (right panel) as a function of $R$ in the galactic plane. The scatter of the dispersion ratios is smaller than the thickness of the lines. As a comparison, dispersion ratios in Solar neighbourhood in the Milky Way (MW) are shown, based on the Hipparcos data \citep[][green markers]{Dehnen:1998} and the SDSS Stripe 82 stars \citep[][black points]{Smith:2012}. For a sample of Sa galaxies, \citet{vanderKruit:1999} found the average dispersion shape using foreknown correlations (dark-green line with error corridor).}
	\label{fig:disp.radial}
\end{figure} 

From the analysis of stellar proper motion data, velocity dispersion ellipsoid parameters have been determined only in the Solar neighbourhood of the Milky Way \citep{Dehnen:1998, Budenbernder:2014}. It is seen from Fig.~\ref{fig:orient} that the velocity ellipsoid orientations inside this ``Solar cylinder'' measured by \citet{Budenbernder:2014} are rather similar with our calculated ellipsoid orientations for M31. In their study of the Solar neighbourhood kinematics, \citet{Dehnen:1998} did not include the vertical dependence of the orientation of the velocity ellipsoid, therefore no quantitative measurements can be made, but qualitatively the results are similar. 

The calculated shape parameters of the velocity dispersion ellipsoids, the ratios $\sigma_z/\sigma_R$ and $\sigma_\theta/\sigma_R$ as a function of galactocentric radius are given in Fig.~\ref{fig:disp.radial}. It is seen that the velocity dispersion ellipsoid for the disc is radially rather elongated, less than in the Milky Way. For the bulge component the ellipsoids are roughly spherical.

\subsection{Comparison of the model with the observed line-of-sight dispersions and velocities}
\label{sec:m31_obs}

\begin{figure}
	\includegraphics[width=84mm]{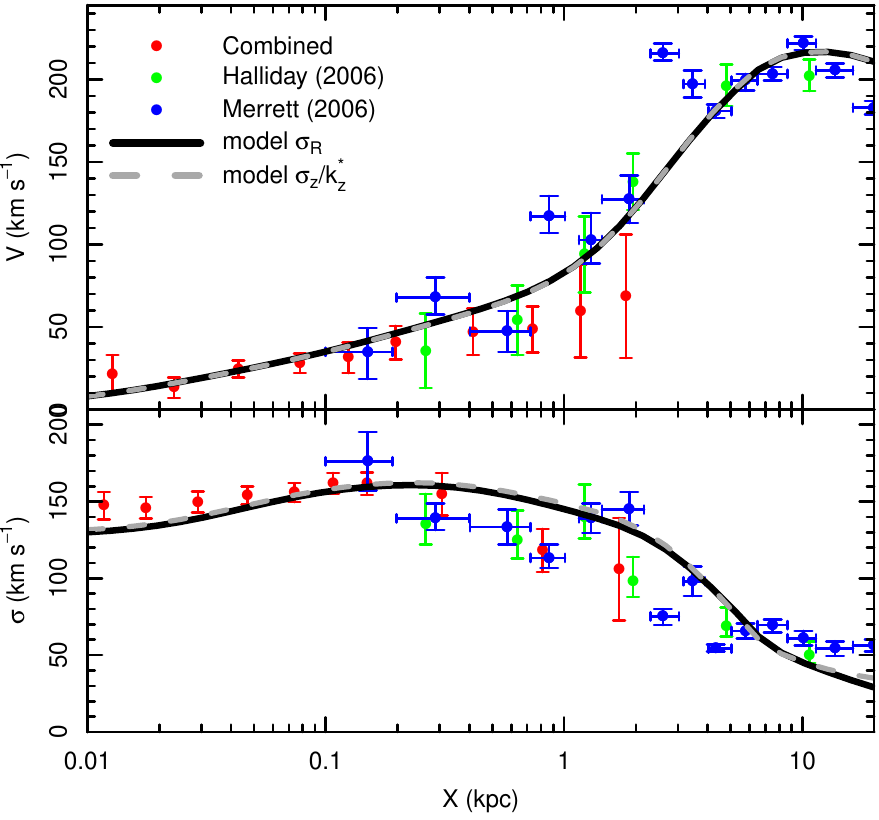}
	\caption{Stellar rotation curve (top panel) and velocity dispersions (bottom panel) along the major axis of the galaxy. Colour points with error bars show the observed kinematics of stars (red points; see text for references) and planetary nebulae \citep{Halliday:2006,Merrett:2006}. The  model curve (based on $\sigma_R$ or $\sigma_z$) is shown as the black/grey line, respectively. In the central and outer regions, the agreement between the model and the observations is not perfect because our model does not include the nucleus component and the stellar halo. 
	}
	\label{fig:rot.curve}
\end{figure}

The ultimate test for every model is a comparison with an experiment or observations. To try our model, we use the stellar kinematics along the major axis of the galaxy measured by \citet{Mcelroy:1983}, \citet{Kormendy:1988}, \citet{vanderMarel:1994}, and \citet{Kormendy:1999}. For noise reduction, we have combined these data as described in \citet{Tempel:2007}. Additionally, we use the observed planetary nebulae kinematics along the major axis as derived by \citet{Halliday:2006} and \citet{Merrett:2006}, and also stellar kinematics off the major axis measured by \citet{Saglia:2010} and \citet{Zou:2011}. 

To make the model comparable with the observations, we integrated the model kinematics over the line of sight and approximated the resulting velocity profile with a normal profile as described in Section~\ref{sec:losvd}.

In Fig.~\ref{fig:rot.curve} we show the observed rotational velocities and velocity dispersions along the major axis of the galaxy. The upper panel shows that our model agrees very well with the observed rotation curve. Since the resultant rotation curve agrees well with the observations across the whole range of radii, the photometry-based division of the galaxy into a bulge and a disc must have been done properly and M31 is indeed dominated by two dynamically different components. The lower panel in Fig.~\ref{fig:rot.curve} shows the observed velocity dispersions and the modelled ones.  In general, the model traces the observations well. We stress that the model is calculated without taking into account the observed dispersions.

\begin{figure}
	\includegraphics[width=\columnwidth]{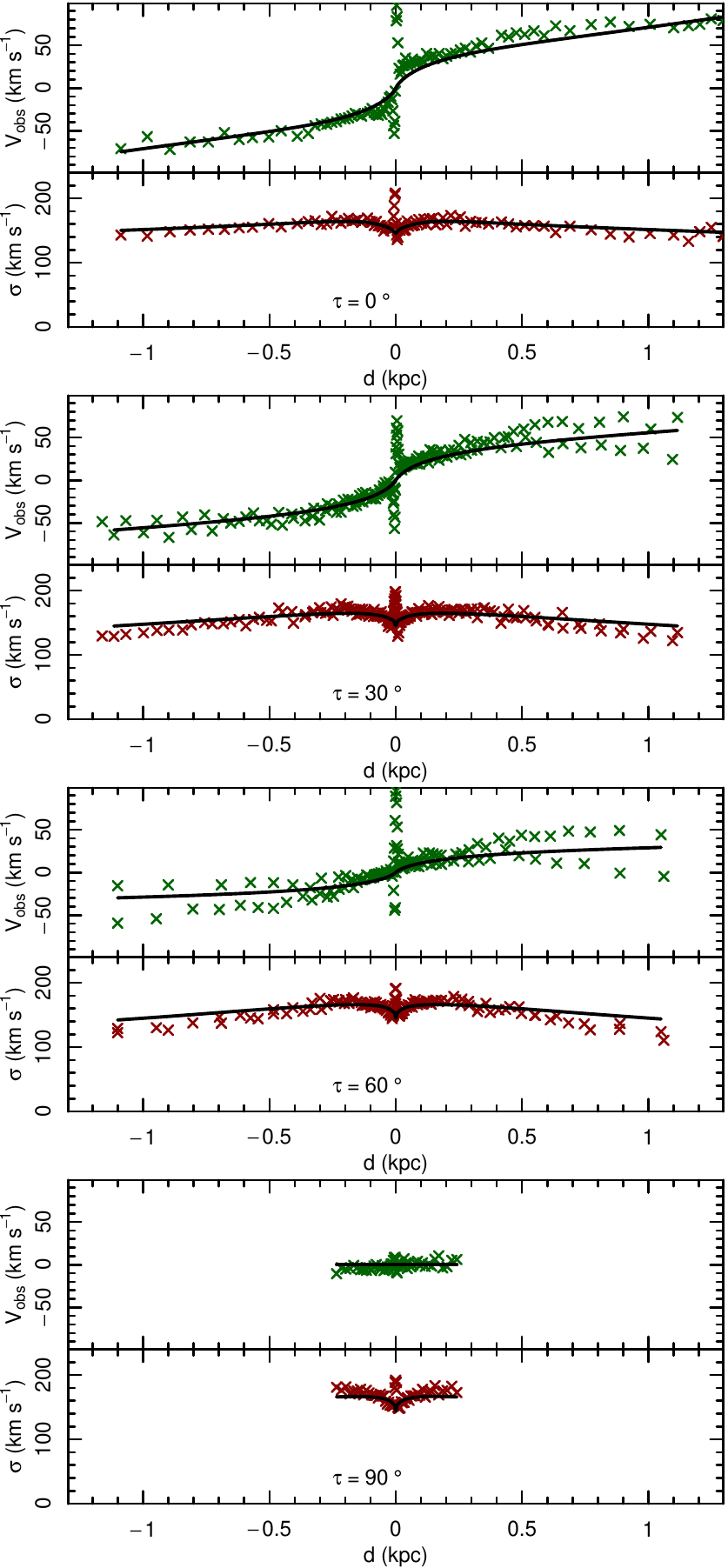}
	\caption{Rotation velocities (top panels) and velocity dispersions (bottom panels) along different slits within the bulge region of M31. The data points represent stellar kinematics as observed by \citet{Saglia:2010} along four slits crossing the galaxy centre, tilted by $0\degr$, $30\degr$, $60\degr$ and $90\degr$ with respect to the major axis of the galaxy. In each panel, $x$-axis shows distance from the galaxy centre along the slit in kiloparsecs. The corresponding model kinematics is shown with the solid line. The mismatch at the central part comes from the nucleus component, not taken into account in our model. }
	\label{fig:saglia}
\end{figure}

One advantage of our model is that it can also be used to calculate stellar kinematics in arbitrary locations within the galaxy, mimicking e.g. observations through a spectroscopic slit intersecting with the major axis. Such measurements of stellar kinematics of the bulge region of M31 along differently tilted slits have been conducted by \citet{Saglia:2010}. Fig.~\ref{fig:saglia} compares these data to our model. The agreement is generally very good; the largest deviations occur along the minor axis ($\tau=90\degr$), where the modelled dispersions remain slightly but systematically lower than the observed ones. This effect is probably caused by the fact that we did not include the nucleus as a separate component in our model.

\begin{figure}
	\includegraphics[width=84mm]{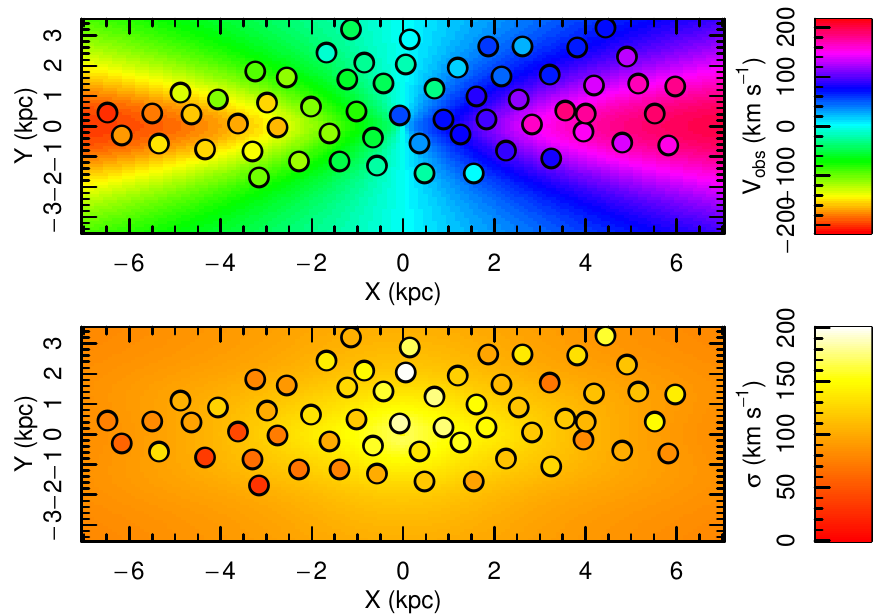}
	\caption{Maps of rotational velocity (top panel) and velocity dispersions (bottom panel) of M31 in the plane of the sky. The smooth background shows the stellar kinematics calculated from our model, the circles correspond to measurements from \citet{Zou:2011}. The observed and modelled velocities, shown with the same colour scheme, match with each other well.}
	\label{fig:zou.grid}
\end{figure}

\citet{Zou:2011} have measured velocities and velocity dispersions in a number of locations across the entire visual galaxy. Fig.~\ref{fig:zou.grid} shows a comparison of this data and our model; once again, the agreement is reassuring. This indicates that the mass distribution model is accurate not only along the major axis, but also across the entire galaxy, and that overall, the third integral of motion is well applicable in practice.

\section{Discussion and conclusions}
\label{sec:conclusions}

In this paper we constructed a dynamical galaxy model in which the kinematics is calculated from the Jeans equations, taking into account the theory of a third integral of motions. We assumed that the integral is in the form described by \citet{Kuzmin:1953, Kuzmin:1956}. The model can be used by fixing the mass distribution of a galaxy on the basis of observed surface brightness distribution and measurements of the gas rotation curve or from stellar spectra and chemical evolution models; the remaining free parameters can be fitted so that the consistency of the Jeans equations with the third integral is highest. If a satisfactory consistency cannot be achieved (as in the case of M31 disc), there can be two explanations: either the integral is not suitable for the given density distribution, or one model assumption(s) is(are) not strictly valid. As an example, a galaxy may be only in a quasi-equilibrium state. In these cases, one can use the solution as an approximation, but with a precaution. It is possible to improve the quality of the consistency by adopting a better relation between the circular and rotational velocity, or by relaxing the strictness of the third integral to a quasi-integral. 

Finally, we integrated the velocity distribution over the line of sight to derive the kinematics (stellar rotational velocities and velocity dispersions) that are directly comparable with observations. The model was built in a general way that allows to calculate the kinematics at any location within the galaxy.

One feature of the model is the possibility to estimate the shape and inclination of the velocity dispersion ellipsoid in the framework of the third integral theory. In general, the third integral can be solved directly only near to the plane of a galaxy\footnote{One exception is the St\"ackel potential, where the third integral is analytical, but the form is very restrictive for practical use \citep{Binney:2008}.}, hence the presented model also extends the usage of the third integral theory further, to regions off the galaxy plane.

We tested our model on the nearby galaxy M31, taking the density distribution from \citet{Tamm:2012} and constraining the model using the observed stellar rotation curve along the major axis. The calculated stellar rotation and velocity dispersions reproduce the actual measurements across the entire galaxy very well, suggesting that the third integral of motion can be used to model the dynamics of M31. We stress that due to some degeneracies (e.g. $a_2$ is slightly degenerated with $z_0$), the exact model parameters cannot be restored uniquely, but this does not influence the model comparison with observations (see Section~\ref{sec:m31_jeans}). Since M31 is a rather typical disc galaxy, the third integral of motion should be an adequate approximation for disc galaxy kinematics in general. However, this conclusion still needs to be validated, which is a planned work for the future.

The derived parameter $z_0$ (the foci of elliptical coordinates, see Fig.~\ref{fig:alpha}) in Eqs.~(\ref{i3}) or (\ref{i33}) have different values for the bulge and for the disc. Within these components their values remain constant, thus the third integrals are precise integrals in both cases. However, for the galaxy as a whole, i.e. a superposition of the bulge and the disc, no single ellipsoidal coordinate system describing the orientation of the velocity dispersion ellipsoids exists -- the coordinate system takes a more complicated form. The weighted average of the parameter $\bar{z_0}$ is constant in regions where either the bulge or the disc dominates, but is a function of coordinates $\bar{z_0}(R,z)$ in the transition region. Thus, for a galaxy as a whole the third integral is actually a quasi-integral. 

We saw from Fig.~\ref{fig:disp.match} (right panel, blue lines) that $\sigma_R$ calculated from two Jeans equations do not match exactly and disagreement increases with $z$. One way to explain the inconsistency between two equations is to accept that a galaxy is in  a quasi-equilibrium state and secular evolution due to fluctuating part of the gravitational potential (irregular forces). An assumption in Jeans equations is that there is no systematic motion in $R$ and $z$ directions. However, when studying the secular evolution of a stellar system due to irregular forces \citep{Kuzmin:1963} derived that in addition to other effects irregular forces cause certain systematic motion in $z$ direction being proportional to $z$.  We intend to study this possibility in the future.

We can also draw some conclusions about the velocity ellipsoid in general, relying on M31 as a typical spiral galaxy with its distinct bulge and disc regions. The velocity ellipsoid is approximately isotropic in the central parts (see Fig.~\ref{fig:disp.radial}), which allows to use simpler models to describe the dynamics in the bulge region. In the outer regions where disc dynamics dominates, the velocity ellipsoid flattens slightly in the $z$-direction.

In our model, the orientation of the velocity ellipsoid is determined by the third integral of motion, which requires that the ellipsoid is oriented along the elliptical coordinate axes (see Fig.~\ref{fig:orient}). Alternatively, the velocity ellipsoid can be assumed to be aligned with the cylindrical coordinate axes, as used in \citet{Cappellari:2008}. According to our model (see Fig.~\ref{fig:orient}), the latter assumption is a good approximation only in the central part of the galaxy, where the ellipsoid is roughly spherical and does not have any distinctive orientation.

The full advantage of the presented model can be taken if applied to large integral field spectroscopic surveys. Several such surveys are in progress or planned, for example SAURON  \citep{deZeeuw:2002}, CALIFA \citep{Sanchez.small:2012}, MANGA \citep{Bundy.small:2015},  TKRS2 \citep{Wirth:2015} and Hector \citep{BlandHawthorn:2015}. We plan to adapt our model to apply it to such larger surveys in the foreseeable future, opening a good opportunity to study the dynamics of a representative set of galaxies and to validate the broader applicability of the third integral of motion.

\section*{Acknowledgments}
We acknowledge the support by the Estonian Research Council grants IUT26-2, IUT40-2, IUT2-27, and by the European Regional Development Fund (TK133). This research has made use of the NASA’s Astrophysics Data System Bibliographic Services. We also thank the referee for helping to significantly improve our paper. 


\begin{appendix}
\section{Solving the Jeans equations and calculation of characteristic curves}\label{sec:the_solving}
\begin{figure*}
		\includegraphics[width=\textwidth]{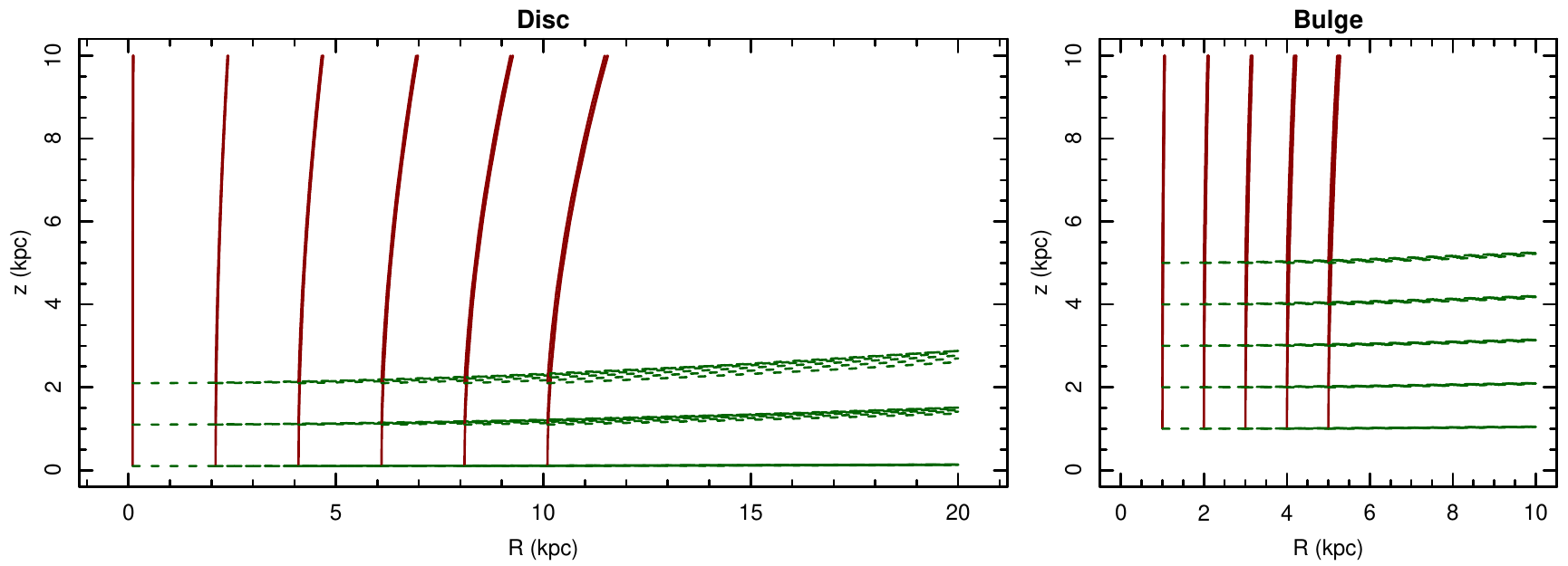}
		\caption{Characteristic curves for the disc (left panel) and bulge (right panel) components. Green dashed lines are characteristic curves for the first Jeans equation calculated from Eq.~(\ref{eq:characteristic_kappa}), red solid lines are characteristic curves for the second Jeans equation calculated from Eq. (\ref{eq:characteristic_ksii}).}
		\label{fig:characteristics}
\end{figure*}

In essence the Jeans equation (\ref{Jeanseq1}) is an equation in a form of
\begin{equation}
	\frac{\partial f}{\partial R} + A(R,z)\, f + B(R,z) \frac{\partial f}{\partial z} = C(R,z) ,
	\label{eqA1}
\end{equation}
where $f = \rho \sigma_R^2$ is a function to be calculated. We solve this equation with a method of characteristics. Substituting partial derivative $\partial f /\partial R$ in (\ref{eqA1}) from the expression for the total differential $\frac{df}{dR} = \frac{\partial f}{\partial R} + \frac{\partial f}{\partial z}\frac{dz}{dR}$, we have (\ref{eqA1}) in form of
\begin{equation}
	\frac{df}{dR} + A(R,z)\, f + \left[B(R,z) - \frac{dz}{dR} \right] \frac{\partial f}{\partial z} = C(R,z).
	\label{eqA2}
\end{equation}
This equation reduces to a simple ordinary differential equation 
\begin{equation}
	\frac{df}{dR} + A(R,z)\, f = C(R,z) ,
	\label{eqA3}
\end{equation}
the solving of which must be done by integrating along the characteristic curves given by equation
\begin{equation}
	\frac{dz}{dR} = B(R,z) .
	\label{eqA4}
\end{equation}

Solution of (\ref{eqA3}) can be written in form of (\ref{eq:solve_jeans_sigma_r}) with integration along the characteristics.

Solution of the second Jeans equation (\ref{Jeanseq2}) can be derived in a similar way.
	
We derived the characteristic curves numerically with the fourth order Runge-Kutta method, following equations (now in designations used in Eqs. (\ref{Jeanseq1}) and (\ref{Jeanseq2})) 
\begin{equation}
		\frac{\mathrm{d}z}{\mathrm{d}R} = \kappa^*(R,z)\label{eq:characteristic_kappa}
\end{equation}
for the first Jeans equation, and
\begin{equation}
	\frac{\mathrm{d}R}{\mathrm{d}z} = \xi^*(R,z)\label{eq:characteristic_ksii}
\end{equation}
for the second one.

Shapes of some characteristic curves are given in Fig.~\ref{fig:characteristics}. 	
\end{appendix}

\label{lastpage}
\end{document}